\documentclass[prl,twocolumn,amsmath,amssymb,superscriptaddress,showpacs]{revtex4}

\usepackage{graphicx}
\usepackage{epsfig}
\usepackage{epstopdf}

\begin{document}

\title{Symmetry breaking of the zero energy Landau level in bilayer graphene}

\author{Y. Zhao}

\affiliation{Department of Physics, Columbia University
New York, NY 10027}

\author{P. Cadden-Zimansky}
\affiliation{Department of Physics, Columbia University
New York, NY 10027}
\affiliation{National High Magnetic Field Laboratory, Tallahassee, FL  32310}

\author{Z. Jiang}
\affiliation{School of Physics, Georgia Institute of Technology,
Atlanta, Georgia 30332, USA}

\author{P. Kim}

\affiliation{Department of Physics, Columbia University
New York, NY 10027}

\date{\today}

\begin{abstract}
The quantum Hall effect near the charge neutrality point in
bilayer graphene is investigated in high magnetic fields of up to
35 T using electronic transport measurements. In the high field
regime, the eight-fold degeneracy in the zero energy Landau level
is completely lifted, exhibiting new quantum Hall states
corresponding filling factors $\nu=$0, 1, 2, \& 3.  Measurements
of the activation energy gap in tilted magnetic fields suggest
that the Landau level splitting at the newly formed $\nu=$1, 2, \&
3 filling factors are independent of spin, consistent with the
formation of a quantum Hall ferromagnet. In addition, measurements
taken at the $\nu$ = 0 charge neutral point show that, similar to
single layer graphene, the bilayer becomes insulating at high
fields.
\end{abstract}

 \maketitle

The unique chiral nature of the carrier dynamics in graphene
results in a novel integer quantum Hall (QH) effect that
distinguishes these atomically thin graphitic materials from
conventional 2-dimensional systems ~\cite{Novoselov_QH, Zhang_QH,
Novoselov_BLG}.  The chiral nature of carriers in single layer
graphene (SLG) and bilayer graphene (BLG) result in unevenly
spaced Landau levels (LL) including a distinctive level located
precisely at the particle-hole degenerate zero energy. While the
energy spacing of SLG LLs scales as the square root of the LL
index $n$ and the square root of the field $B$, BLG levels have
energies given by~\cite{Mccann}
\begin{equation}
\epsilon_n^\pm=\pm\hbar\omega\sqrt{n(n-1)},
\end{equation}
where $\omega=eB/m$ is the cyclotron frequency with the BLG band
mass $m\approx0.05\,m_e$, and $n$ is a non-negative integer
representing Landau orbit index in each layer.  While the valley
and spin degrees of freedom lead to a 4-fold degeneracy in
graphene LLs, the additional orbital degeneracy for the $n=$0 \& 1
indices in BLG result in a 8-fold degenerate, zero energy level
that is unprecedented in LL physics ~\cite{Mccann, Novoselov_BLG,
Barlas}.

At high fields, the decreasing radius of the cyclotron orbits
gives rise to increasing electron-electron interactions which can
perturb the degenerate LLs.  Experiments on SLG have demonstrated
this high-field symmetry breaking through the appearance of new QH
states at the filling factor sequences $\nu=0, \pm 1,\pm
4$~\cite{Zhang_HQH, Jiang_HQH, Maan_HQH}. The precise nature of
the field-dependent mechanisms that lift the degeneracies is still
under experimental and theoretical debate ~\cite{Jiang_SSC,
Yang_SSC}. In particular, it has recently been observed that in
SLG the $\nu=0$ QH state becomes increasingly insulating at higher
fields~\cite{Checkelsky, Checkelsky2}.  Similar to SLG, the
enhanced interactions under high magnetic field is expected to
lift the 8-fold zero energy LL degeneracy in BLG. A number of
theoretical predictions involving unusual collective excitations
have been proposed to occur in this particle-hole symmetric LL as
its degeneracy is broken~\cite{Barlas, Shizuya, Abanin,
Nandkishore}, but the experimental observation of this broken
symmetry is yet to be observed.

In this Letter we present transport measurements demonstrating the
hierarchy of the splittings for the zero-energy LL in bilayer
graphene as the external magnetic field is increased. Activation
energy measurements of these splittings demonstrate energy spacing
smaller than the bare electron-electron interaction energy and
Zeeman energy.  Tilted field measurements indicate that the
degeneracy breaking at filling factors above the charge neutral
point are independent of spin, consistent with the formation of a
quantum Hall ferromagnet.  While a quantum Hall plateau in the
transverse resistance is observed at the charge neutral point, the
longitudinal resistance at this point is seen to diverge to an
insulating state with increasing field.

\begin{figure}
\includegraphics[width=1.0\linewidth]{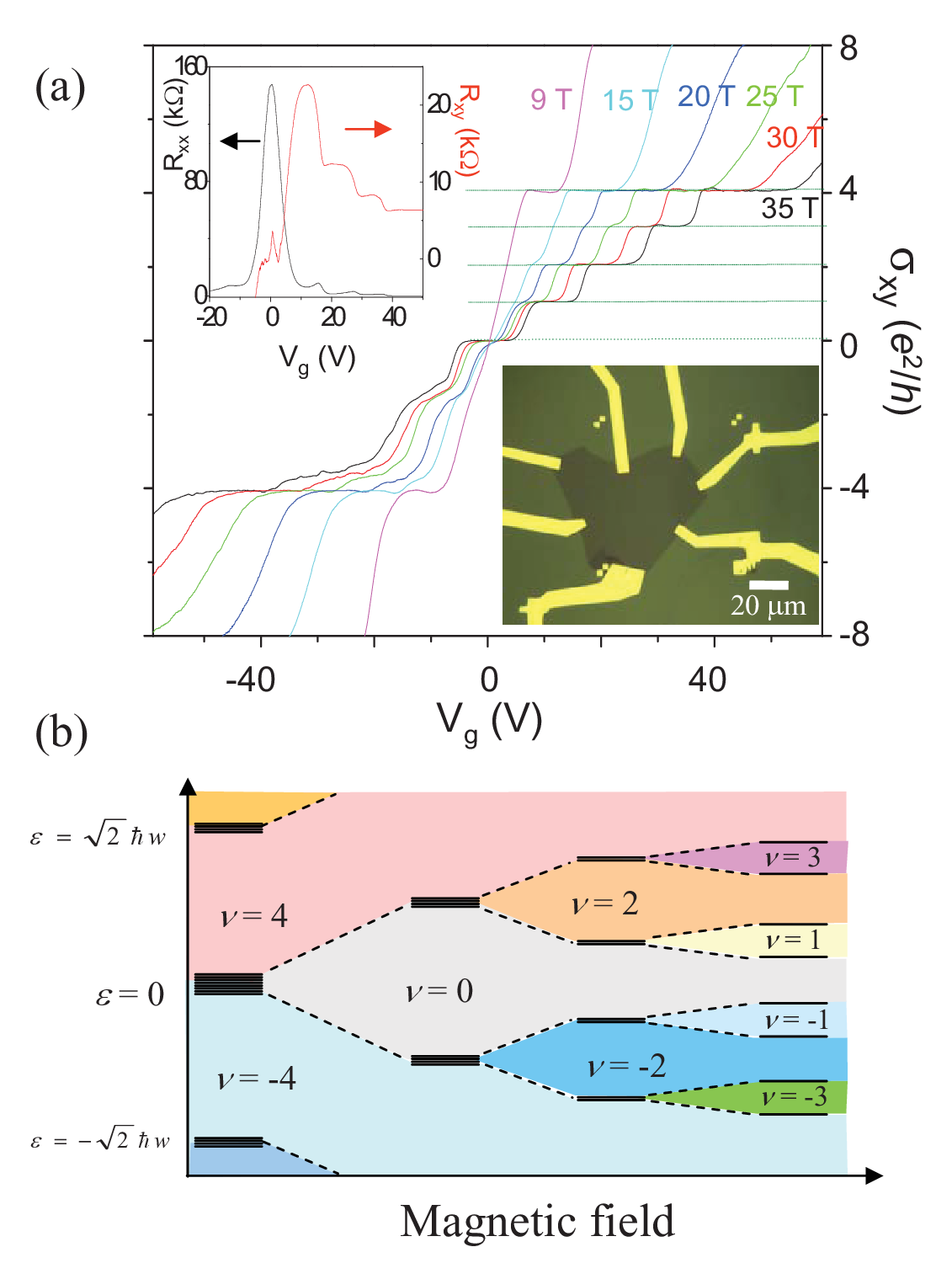}
\caption{(a) Hall conductivity $\sigma_{xy}$, as a function of
gate voltage $V_g$ at T=1.4K at different magnetic fields: 9, 15,
20, 25, 30, and 35~T. Upper inset: $R_{xx}$(in black) and
$R_{xy}$(in red) as $V_g$ varies at B=35~T. Lower inset: optical
microscope image of a BLG device used in this experiment. (b)
Schematic of the zero energy LL hierarchy in bilayer graphene at
high magnetic field.}\label{Fig.1}
\end{figure}

The graphene samples used in this work are deposited on
SiO$_2$(300~nm)/Si substrates by mechanical exfoliation techniques
from bulk single crystals~\cite{Novoselov_PNAS}. The number of
graphene layers are identified by optical contrast, cross-checked
using Raman spectroscopy~\cite{Ferrari} and measurements of the
high-field LL spectrum.  Au/Cr electrodes for the graphene are
fabricated by conventional electron beam lithography followed by
metal evaporation. A gate voltage $V_g$ is applied to the
degenerately doped Si substrate to control the carrier density
$n_c$ according to the relation $n_c=C_g(V_g-V_D)/e$, where the
areal gate capacitance is $C_g=7.1\times
10^{10}\,e/\rm{V}\cdot\rm{cm}^2$ and $V_D$ is the gate voltage
corresponding the charge neutrality point. The lower inset of
Fig.~1a shows an optical microscope image of a typical BLG device
used in this study. This device has a mobility as high as $\sim 1
\times 10^4\,\rm{cm}^2/\rm{V}\cdot\rm{s}$ measured at the carrier
density $n_h=4\times 10^{12}\,\rm{cm}^{-2}$. Since the electrodes
of this device are configured in non-ideal Hall bar geometry, the
magnetoresistance $R_{xx}$ and Hall resistance $R_{xy}$ were
obtained following the van der Pauw method with symmetric
($R_{xx}$) and anti-symmetric ($R_{xy}$) averaging over data from
positive and negative magnetic fields. The Hall conductivity
$\sigma_{xy}$ is then computed from $R_{xx}$ and $R_{xy}$.

Fig.~1(a) shows $R_{xx}$ and $R_{xy}$ (upper inset) and
$\sigma_{xy}$. As the back gate voltage alters the carrier
density, QH plateaus in $R_{xy}$ with corresponding zeros in
$R_{xx}$ are observed. Consequently, the calculated $\sigma_{xy}$
shows well quantized plateau values at
$\frac{1}{\nu}\frac{h}{e^2}$ with the integer filling factor
$\nu$. In the low magnetic field regime ($B<10$~T), QH states
corresponding to $\nu=\pm 4, \pm 8$ are present, as were
previously seen in Ref. \cite{Novoselov_BLG}. As $B$ increases,
however, new QH states emerge, as evidenced by additional QH
plateaus at $\nu=0\,\&\,2$ and then $\nu= 1\,\&\,3$ at higher
fields.

In the high magnetic field regime, $B\gtrsim20$~T, the new set of
QH states demonstrate the full 8-fold degeneracy lifting in the
zero energy LL.  On close inspection of $\sigma_{xy}$ in different
magnetic fields, we can construct the hierarchical evolution of
these new QH states. As the field increases from the low magnetic
field regime, the $\nu=0$ and 2 states start to develop at 15~T
and are fully evolved by 20~T; while the $\nu=1$ and $\nu=3$
states start to develop at 20~T and are fully evolved by 25~T.
This hierarchical appearance of the new QH states is in accordance
with the sequential symmetry breaking of the zero energy LL
degeneracy as depicted in Fig.~1(b) and suggests that different
symmetry-breaking process are relevant as $B$ increases.

\begin{figure}
\includegraphics[width=1.0\linewidth]{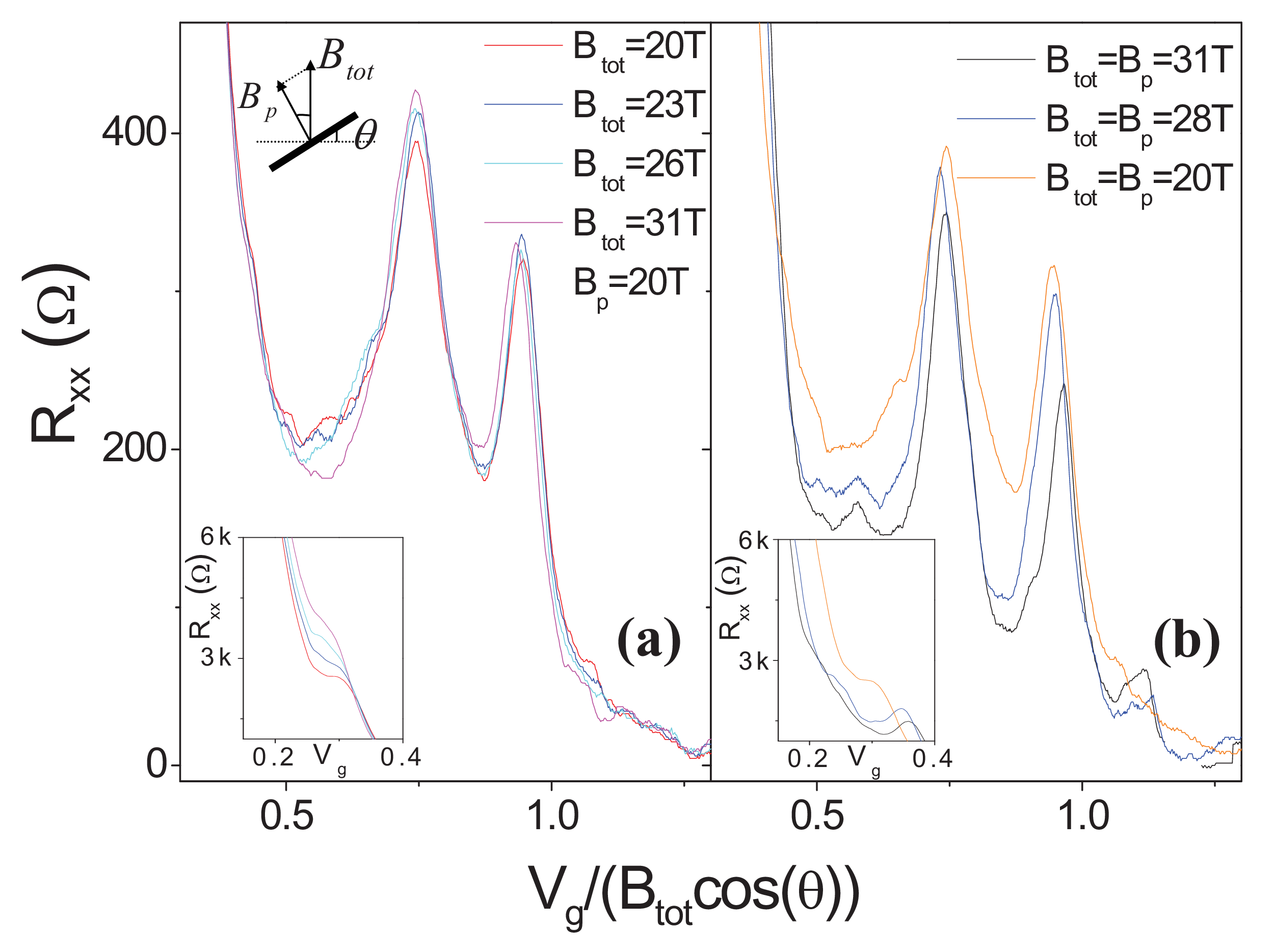}
\caption{(a) Magnetoresistance $R_{xx}$ as a function of
normalized back gate voltage $V_g$/($B_{tot}$$\cos(\theta)$)
around $\nu=2$ and $\nu=3$, at four different total magnetic
fields with the same perpendicular field. Upper inset: Schematic
diagram of tilted field; Lower inset: $R_{xx}$ vs.
$V_g$/($B_{tot}$$\cos(\theta)$) at $\nu=1$ state under the same
condition as in the main figure. (b)$R_{xx}$ as a function of
normalized gate voltage with a zero tilting angle at different
total fields, with the same scale of Fig2(a). Lower inset:
$R_{xx}$ vs. $V_g$/($B_{tot}$$\cos(\theta)$) at $\nu=1$ state. All
data are taken at 1.5~K. }\label{Fig.2}
\end{figure}

Mechanisms for breaking the LL degeneracy include disorder,
lattice strain, and charged impurities.  The magnitude of the last
of these can be estimated by assuming that the offset of the Dirac
point gate voltage $V_D$ from zero energy reflects a bilayer
capacitively charged across the $d = 0.34 nm$ layer spacing.  From
this model our measured $|V_D|\sim2$ V corresponds to a maximum
charging energy of $\sim$ 2 meV, small than, but of the same order
as the high-field (B $\gtrsim$ 20T) bare Zeeman energy.  In
addition to the above mentioned graphene impurities, there are two
field-dependent factors that can lead to the lifting of the eight
degeneracies of this LL:  Zeeman splitting of spin in magnetic
field and electron-electron interactions that increase with field
as the radius of the cyclotron orbits decrease.  Zeeman splitting
is given by $\Delta_z=g\mu_B B$, where $\mu_B$ is the Bohr
magneton, and $g$ is the gyromagnetic factor for the carriers in
BLG.  The Coulomb interactions between electrons are given by
$e^2/{\epsilon}{l_B}$, where $\epsilon\sim4$ is the dielectric
constant for graphene and $l_B=\sqrt{\hbar/eB}$ is the magnetic
length.  In order to produce full 8-fold symmetry breaking, these
mechanisms must lift not only the spin and valley degeneracies,
but the $n = 0\,\&\,1$ LL orbital degeneracy. Barlas~\emph{et al.}
have suggested that the exchange term from Coulomb interactions
between electrons in these degenerate orbital states leads to a QH
ferromagnetic state where the 8-fold symmetry is lifted by this
exchange interaction. A Hund rule-like hierarchical symmetry
breaking in the zero energy LL is predicted with this
exchange-enhanced Zeeman splitting followed by the
spin-independent valley and orbital splitting~\cite{Barlas}.

In order to test out this hierarchical degenerate lifting and the
role of electron-electron interactions, we first note that
sublattice or interlayer splittings that are associated solely
with the Coulomb interactions between electrons localized in LL
orbits should depend only on the normal magnetic field ($B_\bot$)
but not the in-plane field ($B_\|$). Whereas spin-splitting
includes a Zeeman component determined by the total field
$B_{tot}=\sqrt{B_\bot^2+B_\|^2}$. Experimentally, this idea can be
tested out by measuring the samples in a series of tilted fields
where we can examine the $R_{xx}$ minima in different $B_\bot$ and
$B_\|$ by tuning the tilting angle $\theta$ and $B_{tot}$. Fig.~2
displays the change of $R_{xx}$ as a function of the gate voltage
normalized by $B_\bot=B_{tot}\cos\theta$, in two different
experimental conditions.

In Fig.~2(a), we first fix $B_p=20$~T and then vary $B_\|$. For
$\nu=2, 3$ QH states, $R_{xx}^{min}$,  the minima of $R_{xx}$
corresponding to the QH zero magnetoresistance, display less than
$\sim10\%$ of variation, indicating no significant changes are
induced by applying $B_\|$. For comparison, in Fig.~2(b), we show
$R_{xx}$ as a function of $B_\bot$ with fixed $\theta=0$, that is
$B_\|=0$ and $B_\bot=B_{tot}$. In this case $R_{xx}^{min}$ for
$\nu=2$ and 3 increases by $\sim40\%$ as $B_\bot$ decreases from
31~T to 20~T, a factor of 4 larger change than the change of
Fig.~2(a) where $B_{tot}$ decreases from 31~T to 20~T while
$B_\bot$ was kept to 20~T. The fact that $R_{xx}^{min}$ for
$\nu=2$ and 3 shows little to no dependence on the in-plane field
strongly suggests that both these QH states are from non-spin
orgin and thus are due to the breaking of the valley or orbital
degeneracy of the zero energy LL.

As for the $\nu=1$ QH state, the experimental data of its tilted
field dependence, shown in the insets of Fig.~2, displays an
increasing $R_{xx}^{min}$ with either a increasing in-plane field
or a increasing normal field. While this evolution is consistent
neither with an electron-electron or spin origin, it is likely
that the increasing $R_{xx}^{min}$ shown in the inset of Fig.~2(a)
is due to the proximity of this filling factor to the increasingly
insulating behavior of the bilayer at the charge neutrality point,
discussed below. However, given the hierarchy of the degeneracy
breaking, schematically shown in Fig.~1(b), the mechanism
underlying the $\nu=1$ breaking should be the same as that for
$\nu=3$, i.e. not from a spin origin, consistent with theoretical
predications of the formation of a QH ferromagnet in BLG at high
magnetic fields \cite{Barlas, Abanin}.

\begin{figure}
\includegraphics[width=1.0\linewidth]{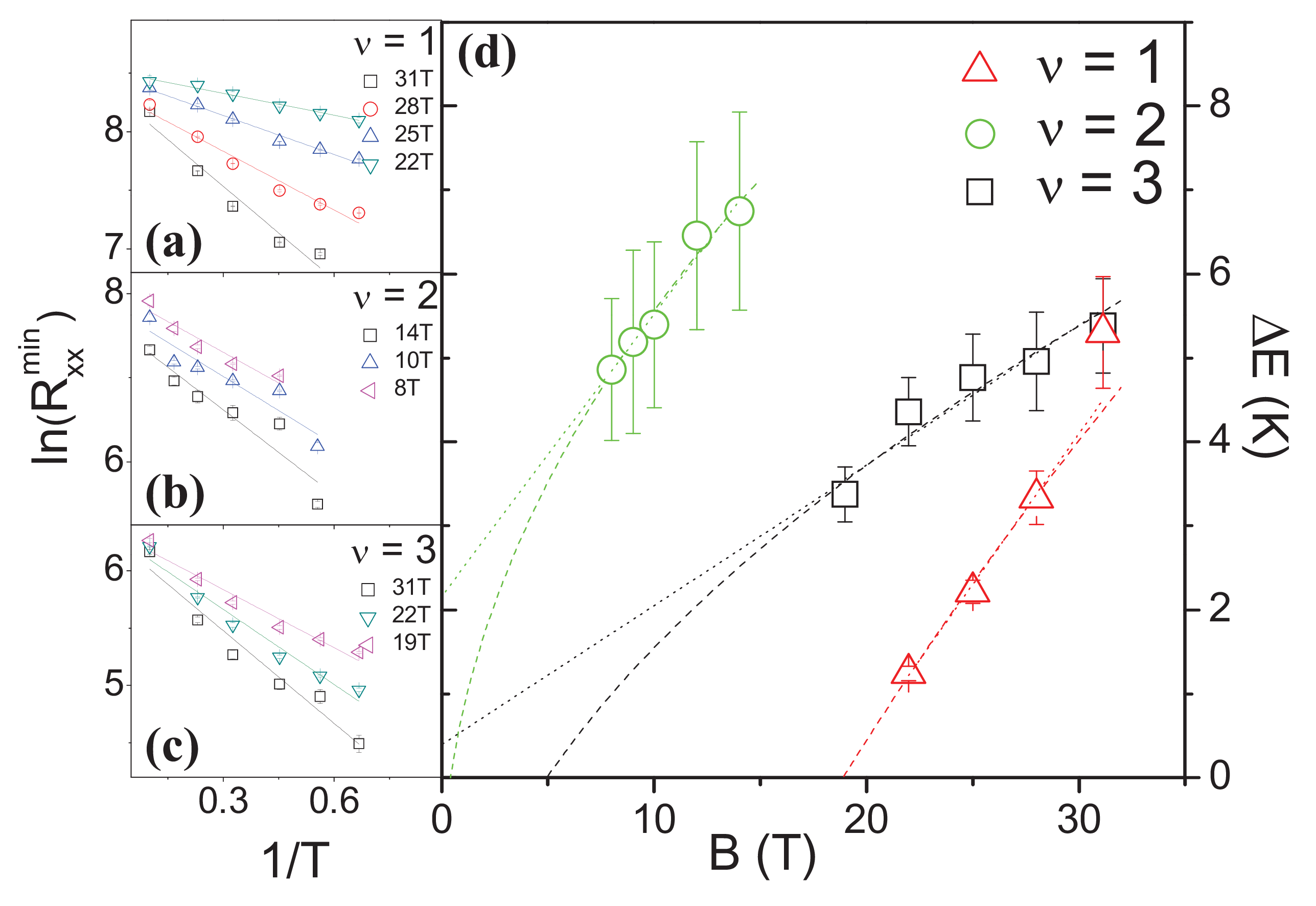}
\caption{(a)Arrhenius plots of $R_{xx}^{min}$'s as a function of
$1/T$ at different fields for $\nu=1$ state, the lines with
respective colors are the linear fits to the data points. (b)
Arrhenius plot for $\nu=2$ state. (c)Arrhenius plot for $\nu=3$
state. (d)Energy gap $\Delta$E vs magnetic field B for different
filling factors $\nu=1$ (open triangle red), $\nu=2$ (open circle
green), $\nu=3$ (open square black). The dotted lines are linear
fits, whereas the dashed lines are square-root fits.}
\label{Fig.3}
\end{figure}

To further understand the nature of the fully lifted LL
degeneracy, we determine the energy of the LL splittings by
measuring $R_{xx}^{min}$ at different temperatures $T$.
Fig.~3(a-c) shows $\log R_{xx}^{min}$ versus $1/T$ for the
$\nu=1,\,2,\,\&\,3$ states. Since $R_{xx}^{min}\propto
\exp{[-(\Delta E-2\Gamma)/2k_BT]}$, where $\Delta E$ is the energy
gap between two subsequent LLs and $\Gamma$ is the LL energy
broadening.  The observed Arrhenius behavior in these plots allows
us to estimate $\Delta E$ at different fields from the slope of
the line fits. Fig.~3(d) displays the field dependence of the
activation gap for $\nu=1,\,2,\,\&\,3$.  Generally, $\Delta E$
increases with increasing $B$ as expected. It is also noted that
at given $B$, $\Delta E_{\nu=2}>\Delta E_{\nu=3}$, $\Delta
E_{\nu=1}$, indicating that the even $\nu$ states have larger
energy than the odd $\nu$ QH states in accordance with the LL
symmetry breaking hierarchy of Fig.~1(b). As for the field
dependence, we find that a $\sqrt{B}$ fit is better for $\nu=2$
and $\nu=3$ states. Attempts to fit the gap evolution linearly to
$B$ result in a positive y-axis intercept for the $\nu=2$ and
$\nu=3$ gaps and a thus a non-physical negative LL broadening
$\Gamma$. From the y-intercept of the $\sqrt{B}$ dependence, we
can extract a physically reasonable LL broadening of
$\Gamma/k_B=1.4$ and 3.7~K for the $\nu=2$ and 3 states,
respectively~\cite{footnote2}. Although the apparent
$\sqrt{B}$-dependence provides an independent confirmation of
non-spin origin of the QH states $\nu=1,2,3$, we note that the
observed energy scale $\Delta E$ is still too small compared to
the Coulomb energy or even the bare Zeeman energy. For $B=15$~T,
the Coulomb energy is 370~K and the Zeeman energy $E_z=21$~K, both
larger than $\Delta E_{\nu=2}=8.6\,\rm{K}$ at this field. It is
possible that large amounts of disorder near the charge neutrality
point of BLG are responsible for such a reduced transport gap.

\begin{figure}
\includegraphics[width=1.0\linewidth]{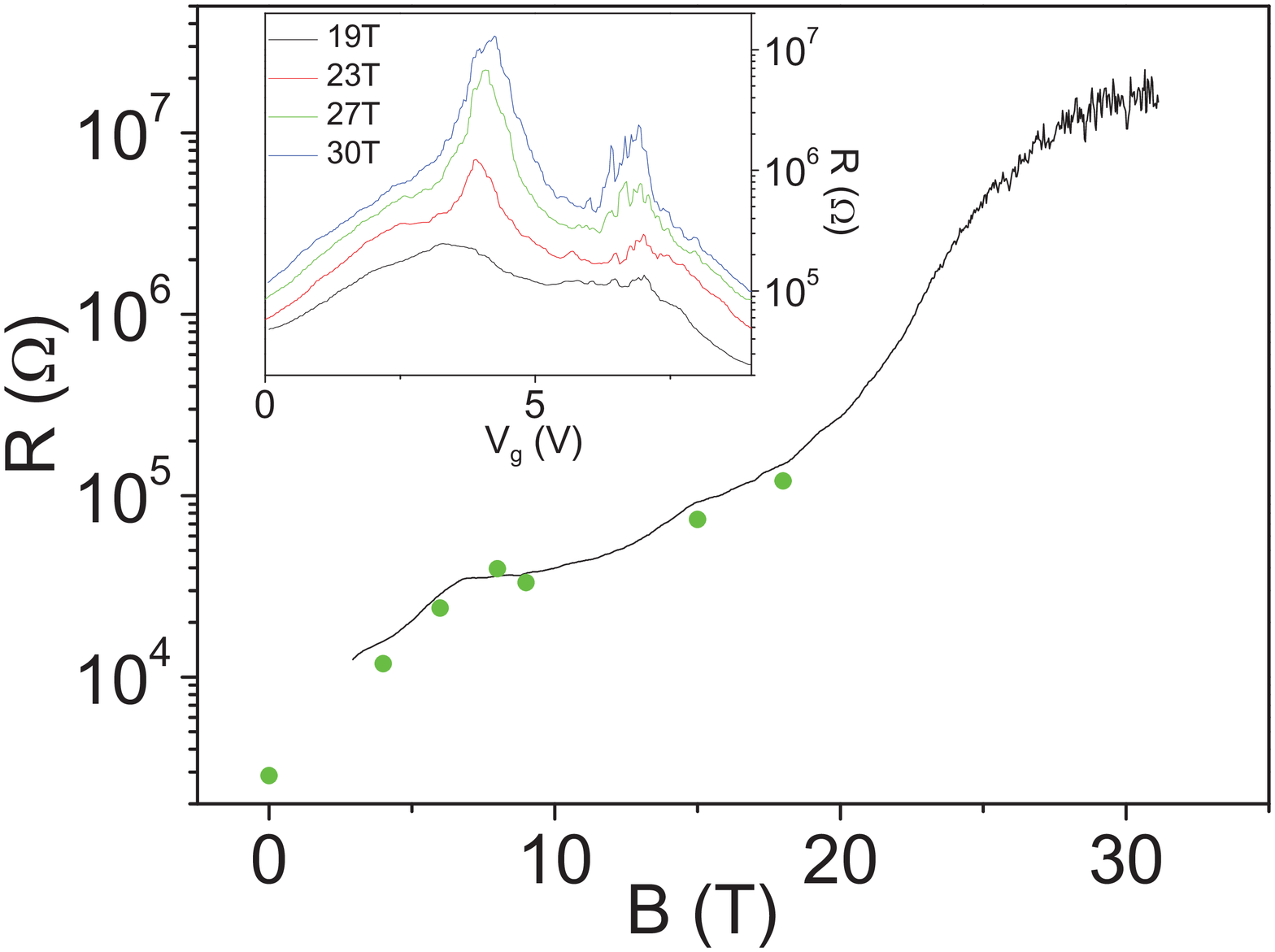}
\caption{Maximum resistance, measured in a two-probe constant
voltage bias method, as a function of magnetic field, which is
applied normally to the graphene plane, at a fixed gate voltage
$V_g$=4.1V around the charge neutrality point. The green dots are
$R_{xx}^{max}$'s at $\nu=0$ using a four-probe measurement. They
are taken at lower magnetic fields($B<20$~T), and multiplied by a
factor of 4 to match the 2 probe data. Upper left inset: Gatesweep
of two-probe resistance at different fields: 19, 23, 27, and 29~T
from bottom to top. All data are taken at 1.5K.} \label{Fig.4}
\end{figure}

We finally focus our attention on the $\nu=0$ QH state. As with
SLG, the presumed $\nu=0$ splitting at the charge neutral point is
not directly observable as a zero in $R_{xx}$. Rather, $R_{xx}$
displays a maximum at this point, whose value increases with
increasing $B$. In Fig.~4, we display our measurement of $R_{xx}$
at the charge neutrality point in BLG. The measured resistance
shows quasi-exponential growth as $B$ increases, up to
$\sim10$~M$\Omega$ at 30~T. To avoid the self-heating of the
graphene discussed in Ref. \cite{Checkelsky2} and to measure
resistances $>$ 10 M$\Omega$ we employ a 2-probe AC measurement
configuration with a constant voltage bias of 500 $\mu$V,
resulting in only $\sim$ 10 fW of heating at the highest fields.
The contact resistance included in this measurement set-up is
relatively small ($\sim$ 1 K$\Omega$), but as an additional
confirmation that it does not affect the behavior of the measured
resistance, we cross-checked the 2-probe measurement using a
conventional, current biased, 4-probe measurement at fields low
enough that the resistance could be reliably measured with the
voltage probes input to an amplifier with 10 M$\Omega$ input
impedance. The exponentially divergent $R_{xx}$ behavior at high
magnetic field is similar to analogous measurements that have been
performed on SLG~\cite{Checkelsky, Checkelsky2}, where a
field-induced QH insulator has been proposed. We also note that
the gate sweep (inset to Fig.~4) displays a growing number of
local maximum in $R_{xx}$, presumably due to the inhomogeneous
distribution of these insulating states at high magnetic
field~\cite{DasSarma}.

In conclusion, we have observed the full degeneracy lifting of the
zero energy LL in bilayer graphene.  Independent measurements of
the longitudinal resistance zeros for the newly observed filling
factors as a function of perpendicular field and temperature each
indicate that the degeneracy lifting for the $\nu=1,2, \,\&\, 3$
splittings originate from electron-electron interactions. The
field dependence of the longitudinal resistance at the $\nu=0$
charge neutral point reveals insulating behavior similar in
character to that of single layer graphene.

The authors thank D. Abanin, Y. Zhang, E. A. Henriksen, F.
Ghahari, and Y. Barlas for helpful discussion, and thank S. T.
Hannahs, E. C. Palm, and T. P. Murphy for their experimental
assistance. This work is supported by DOE (No. DEFG02-05ER46215).
A portion of this work was performed at the National High Magnetic
Field Laboratory, which is supported by NSF Cooperative Agreement
No. DMR-0654118, by the State of Florida, and by the DOE.

{\it Note added.}-During the preparation of this manuscript, we
became aware of related work with a similar conclusion from
Feldman {\it et al.}~\cite{Yacoby}.


\begin{thebibliography}{text}
\bibitem{Novoselov_QH}K. S. Novoselov et al., Nature {\bf 438}, 197 (2005).
\bibitem{Zhang_QH}Y. Zhang, Y. Tan, H. L. Stormer, and P. Kim, Nature {\bf 438}, 201 (2005).
\bibitem{Novoselov_BLG}K. S. Novoselov et al., Nature Phys. {\bf 2}, 177 (2006).
\bibitem{Mccann}Edward McCann and Vladimir I. Fal'ko, Phys. Rev. Lett. {\bf 96}, 086805 (2006).
\bibitem{Barlas}Yafis Barlas, R. C\^{o}t\'{e}, K. Nomura, and A. H. MacDonald, Phys. Rev. Lett. {\bf 101}, 097601 (2008).
\bibitem{Zhang_HQH}Y. Zhang, Z. Jiang, J. P. Small, M. S. Purewal, Y. -W. Tan, M. Fazlollahi,
J. D. Chudow, J. A. Jaszczak, H. L. Stormer, and P. Kim, Phys.
Rev. Lett. {\bf 96}, 136806 (2006).
\bibitem{Jiang_HQH}Z. Jiang, Y. Zhang, H. L. Stormer, and P. Kim ,Phys. Rev. Lett. {\bf 99}, 106802 (2007).
\bibitem{Maan_HQH} A. J. M. Giesbers, L. A. Ponomarenko, K. S. Novoselov, A. K. Geim, M. I. Katsnelson, J. C. Maan, U. Zeitler, arXiv:0904.0948).
\bibitem{Jiang_SSC}Z. Jiang, Y. Zhang, Y. -W. Tan, H. L. Stormer, and P. Kim, Solid State Commun {\bf 143}, 14 (2007).
\bibitem{Yang_SSC}K. Yang, Solid State Commun.{\bf 143}, 27 (2007).
\bibitem{Checkelsky}J. G. Checkelsky, L. Li, and N. P. Ong, Phys. Rev. Lett. {\bf 100}, 206801 (2008).
\bibitem{Checkelsky2}J. G. Checkelsky, L. Li, and N. P. Ong, Phys. Rev. B {\bf 79}, 115434
(2009).
\bibitem{Shizuya}K. Shizuya, Phys. Rev. B {\bf 79}, 165402 (2009).
\bibitem{Abanin} D. A. Abanin, S. A. Parameswaran, and S. L. Sondhi, arXiv:0904.0040.
\bibitem{Nandkishore}R. Nandkishore and L. Levitov, arXiv:0907:5395v1
\bibitem{Novoselov_PNAS}K. S. Novoselov, D. Jiang, T. Booth, V. V. Khotkevich, S. M. Morozov,
and A. K. Geim, Proc. Natl. Acad. Sci. U.S.A. {\bf 102}, 10451
(2005).
\bibitem{Ferrari}A.C. Ferrari, J. C. Meyer, V. Scardaci, C. Casiraghi, M. Lazzeri, F. Mauri,
S. Piscanec, D. Jiang, K. S. Novoselov, S. Roth, A. K. Geim, Phys.
Rev. Lett. {\bf 97}, 187401 (2006).
\bibitem{footnote2}For $\nu=1$, the energy gap $\Delta E$ is too
small to make a relable fit for $B$ dependence.
\bibitem{DasSarma}S. Das Sarma, and Kun Yang, arXiv:0906.2209.
\bibitem{Yacoby} B. Feldman, J. Martin, and A. Yacoby, arXiv:0907:5395v1.

\end{thebibliography}
\end{document}